\documentclass[aps,prb,twocolumn,superscriptaddress,10pt]{revtex4-1}
\setcitestyle{numbers,square}
\pdfoutput=1
\usepackage{amsmath,amssymb,amsfonts,upgreek}
\usepackage{color}
\usepackage{graphicx}

\newcommand{\svo}{SrVO$_3$}
\newcommand{\cvo}{CaVO$_3$}
\newcommand{\sto}{SrTiO$_3$}
\newcommand{\lao}{LaAlO$_3$}
\newcommand{\atet}{\ensuremath{a_{\mathrm{tetr}}}}
\newcommand{\ttg}{t$_{\mathrm{2g}}$}
\newcommand{\eg}{e$_{\mathrm{g}}$}
\newcommand{\umit}{\ensuremath{U_{\mathrm{MIT}}}}

\graphicspath{{./}}

\begin{document}

\title{Multilayer engineering of \cvo{} thin films with \sto{} and \lao{} from DFT+DMFT}

\author{Sophie Beck}
\email{sophie.beck@mat.ethz.ch}
\affiliation{Materials Theory, ETH Z\"u{}rich, Wolfgang-Pauli-Strasse 27, 8093 Z\"u{}rich, Switzerland}
\author{Claude Ederer}
\email{claude.ederer@mat.ethz.ch}
\affiliation{Materials Theory, ETH Z\"u{}rich, Wolfgang-Pauli-Strasse 27, 8093 Z\"u{}rich, Switzerland}

\date{\today}

%%%%%%%%%%%%%%%%%%%%%%%%%%%%%%%%%%%%%%%%%%%%%%%%%%%%%%%%

\begin{abstract}
In this paper we use density functional theory combined with dynamical mean-field theory (DFT+DMFT) to study interface effects between thin films of the correlated metal \cvo{} and the two typical substrate materials \sto{} and \lao{}.
We find that the \cvo{}/\sto{} interface has only a marginal influence on the \cvo{} thin film, with the dominant effect being the (bulklike) epitaxial strain imposed by the large lattice mismatch, rendering the \cvo{} film insulating due to the enhanced orbital polarization related to the strong level splitting between the \ttg{} orbitals.
In contrast, at the polar \cvo{}/\lao{} interface, the presence of the interface can have a huge effect on the thin film properties, depending both on the specific interface termination as well as the specific boundary conditions imposed by the multilayer geometry.
We compare three different approaches to model the interface between the correlated metal \cvo{} and the band insulator \lao{}, which all impose a different set of (electrostatic) boundary conditions on the electronic structure.
The spectral properties obtained from our calculations reveal a strong influence of the supercell geometry, ranging from bulklike to highly doped and structurally distorted phases, indicating a potential tunability of the interfacial properties via multilayer engineering.
\end{abstract}

\maketitle

%%%%%%%%%%%%%%%%%%%%%%%%%%%%%%%%%%%%%%%%%%%%%%%%%%%%%%%%
\section{Introduction}

Recent advances in the exploration of oxide thin films and heterostructures have demonstrated a wide range of possibilities in tailoring materials properties by the choice of substrate, film thickness, superlattice periodicity, or growth conditions (e.g., variations in oxygen pressure).
However, not only is it possible to tune the existing properties of the corresponding materials, but also completely new phases, not observed in the corresponding bulk systems, can emerge~\cite{Mannhart/Schlom:2010,Hwang_et_al:2012,Chakhalian_et_al:2014}. 
Interesting examples for the latter are the correlated metals \svo{} and \cvo{}, for which a metal-insulator transition (MIT) under decreasing film thickness has been reported in ultra-thin films~\cite{Yoshimatsu_et_al:2010,Gu_et_al:2013,Gu/Wolf/Lu:2014,Zhong_et_al:2015,Beck_et_al:2018,McNally_et_al:2019}.
Both materials exhibit a perosvkite structure, with a $d^1$ electron configuration of the V$^{4+}$ cation, but \cvo{} is generally assumed to be closer to the MIT, due to its narrower bandwidth and a small crystal-field splitting related to octahedral rotations~\cite{Yoshida_et_al:2010,Pavarini_et_al:2004}.
This results in a particularly high tunability of \cvo{} with respect to temperature, strain, and film thickness~\cite{Beck_et_al:2018,McNally_et_al:2019}.
Both \svo{} and \cvo{} have also been suggested as promising candidates for applications as transparent conductors~\cite{Zhang_et_al:2016}.

Recent computational work has ascribed the MIT in \cvo{} thin films to an interplay of strain and a surface-related crystal-field splitting that favors a Mott-insulating state~\cite{Beck_et_al:2018}.
While both phenomena can in principle be simulated detached from the presence of the substrate, many of the fascinating emerging phenomena observed in functional oxides are specifically due to interface-related effects.
This is also highlighted by the paradigmatic case of the \lao{}/\sto{} interface, that, given the right set of boundary conditions, can exhibit a two dimensional electron gas (2DEG)~\cite{Ohtomo/Hwang:2004}, novel magnetic properties~\cite{Brinkman_et_al:2007}, or even superconductivity~\cite{Reyren_et_al:2007}, despite the fact the both compounds are conventional nonmagnetic band insulators. While the polar discontinuity at the \lao{}/\sto{} interface appears to be crucial, the fact that the 2DEG only forms at the electron-doped $n$-type interface, i.e., the (LaO)$^+$/(TiO$_2$)$^0$ interface, indicates that the emerging interface phenomena in oxide heterostructures generally result from a complex interplay between structural (e.g., strain, octahedral rotations, cation intermixing, or stoichiometry) and electronic (e.g., charge transfer and band alignment) interfacial reconstruction mechanisms.
Due to the large number of compensation mechanisms at play, and since many of these effects are difficult to control and isolate both in experiment and in theory, their relative importance  is still controversially discussed~\cite{Nakagawa/Hwang/Muller:2006,Yamamoto_et_al:2011}.

In view of this, it is apparent that a comprehensive understanding of the thickness-induced MIT in \cvo{} thin films also requires to address potential effects originating from the film/substrate interface.
Here, we study such effects by explicitly considering two commonly used substrate materials, \sto{} and \lao{}.
In order to accurately describe the specific chemical environment and all structural effects, and to simultaneously include the dynamical correlation effects responsible for the metal-insulator transition in the absence of any magnetic or other symmetry-breaking long-range order, we use a combination of density functional theory plus dynamical mean-field theory (DFT+DMFT).

To this end, we first analyze interface-related changes on the structural level, in particular the evolution of the GdFeO$_3$-type distortion of the octahedral network, which directly affects the bandwidth and induces a potentially relevant crystal-field splitting.
The control of octahedral tilts and rotations via structural connectivity across an interface has recently been suggested as a possible design tool in oxide heterostructures~\cite{Rondinelli/May/Freeland:2012}.
We then address layer-dependent changes in the occupation of the \ttg{} levels of the V$^{4+}$ cation resulting from interfacial charge transfer or from electronic reconstruction due to the polarity of the \cvo{}/\lao{} interface.
In the latter case, we also consider different multilayer geometries to model the \cvo{}/\lao{} interface within different (electrostatic) boundary conditions.
Concomitant to the structural modifications, the polar interfaces can lead to electron- and hole-doped layers of \cvo{} that give rise to vastly different behaviours and an increased metallicity as compared to bulk, which can also compete with other factors such as, e.g., strain and finite size effects.

This paper is organized as follows:
In Sec.~\ref{subsec:DFT} we introduce the different supercell geometries that are used to simulate the interfaces and discuss the resulting boundary conditions, while in Sec.~\ref{subsec:DMFT} we describe the computational details of our DFT+DMFT calculations.
We then first present our results obtained for \cvo{}/\svo{} heterostructures in Sec.~\ref{subsec:CS} before discussing the more complex case of \cvo/\lao{} in Sec.~\ref{subsec:CL}. In both cases we first discuss results obtained on the DFT level, i.e. structural relaxation of octahedral tilts and rotations and initial charge transfer, before we address the corresponding implications on the metal-insulator transition obtained within DMFT.
Finally, we summarize our results and discuss some conclusions in Sec.~\ref{sec:sum}.

%%%%%%%%%%%%%%%%%%%%%%%%%%%%%%%%%%%%%%%%%%%%%%%%%%%%%%%%
\section{Computational method}
\label{sec:cm}

\subsection{Supercell construction}
\label{subsec:DFT}

To model the substrate-film interface within periodic boundary conditions, we are using multilayer-slab geometries corresponding to a stacking of perovskite units along the (pseudo-) cubic $[001]$ direction, which defines the $c$-axis of the unit cell.
Relevant structural aspects to consider in the construction of the corresponding supercells are the epitaxial constraints and the imposed strain due to the lattice mismatch between substrate and thin film material, the octahedral tilt system, the interface termination, and the overall stoichiometry and polarity of the slab, which is discussed in more detail below.

\paragraph{Supercell geometry}
The perovskite structure can be viewed as a periodic stacking of alternating $A$O and $B$O$_2$ sublayers along the $[001]$ direction. Depending on the formal valences of the $A$ and $B$ cations, these sublayers are either charge-neutral, as in ``$2+$/$4+$'' perovskites such as Ca$^{2+}$V$^{4+}$O$_3$ or Sr$^{2+}$Ti$^{4+}$O$_3$, or carry formal charges of $+1$ and $-1$, respectively, as in ``$3+$/$3+$'' perovskites such as La$^{3+}$Al$^{3+}$O$_3$. 
Thus, if a finite number of full unit cells of a $3+$/$3+$ perovskite are stacked along $[001]$, the system gains a polarity, which can be compensated by interface charges, depending on the specific boundary conditions~\cite{Higuchi/Hwang:2012}.
On the other hand, if a slab with symmetric interface terminations is constructed, by adding either an extra $A$O or $B$O$_2$ layer, then the system would be charged if all ions would assume the same valence as in the bulk material. Since charge neutrality is generally enforced within a DFT calculation, this in turn means that in this case not every ion can assume its bulklike valence. 
We note that both cases, the polar and the symmetrically terminated slab, involve a polar discontinuity at the interface between the $3+$/$3+$ perovskite and a material with charge neutral layers (or vacuum), but represent different (electrostatic) boundary conditions.

\begin{figure}
    \centering
    \includegraphics[width=\linewidth]{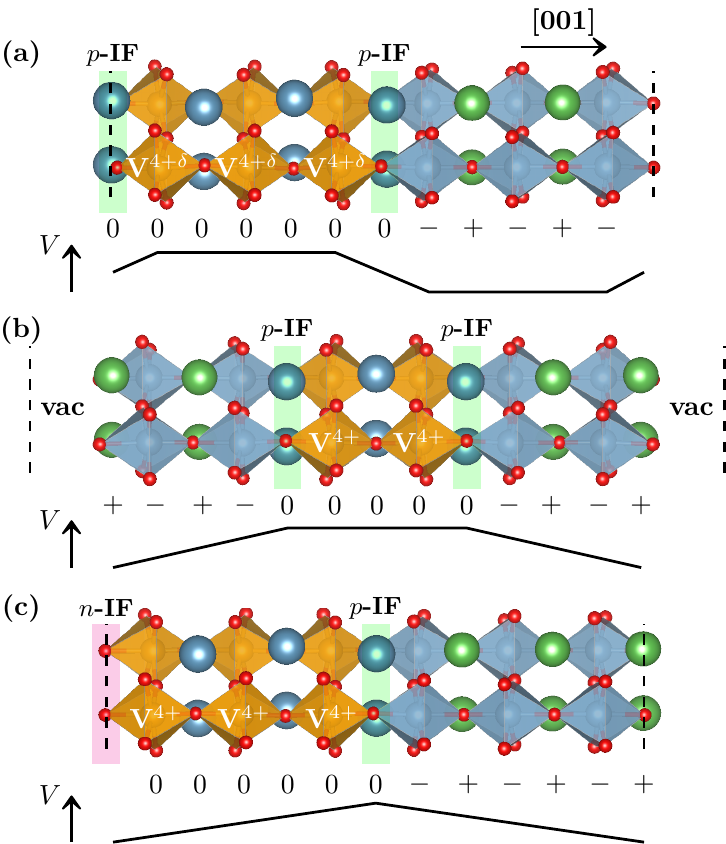}
    \caption{%
    Simplified schematic depiction of the three types of multilayer geometries used to model the CaVO$_3$/LaAlO$_3$ interface.
    Ca atoms are shown in blue, La in green, V in orange, Al in blue, and O in red.
    (a) Symmetric multilayer, with two symmetric (in this case $p$-type) interfaces.
    (b) Symmetric vacuum slab, with two symmetric (in this case $p$-type) interfaces and vacuum separating the \lao{} layers.
    (c) Asymmetric multilayer with two different interfaces.
    The corresponding evolution of the electrostatic potential $V$ is sketched below each multilayer.}
    \label{fig:slabs}
\end{figure}

Thus, in order to compare different scenarios for the case of the CaVO$_3$/LaAlO$_3$ interface, we use three different supercell types, depicted in Fig.~\ref{fig:slabs}.
For the case of the CaVO$_3$/SrTiO$_3$ interface only one type is used, since this interface does not involve a polar discontinuity. 
In all cases, we use the notation ``$i$/$j$'' to specify the thickness of the individual components, where $i$ specifies the number of VO$_2$ sublayers and $j$ the number of $B$O$_2$ sublayers with $B$ either Ti or Al. The corresponding number of CaO and SrO or LaO layers is then uniquely determined by the specific multilayer geometry and the interface termination (see below).

The first supercell geometry, shown in Fig.~\ref{fig:slabs}a), corresponds to symmetric multilayers with identical termination at both interfaces, resulting in either the so-called ``$n$-type'' (VO$_2$)$^0$/(LaO)$^+$ or ``$p$-type'' (CaO)$^0$/(AlO$_2$)$^-$ interface. Here, the names indicate the type of charge that is needed to compensate the (bulklike) formal charges.
Note that in this case the supercell contains ``half unit cells'' of each material. Thus, for example in the case of the $p$-type interface, the resulting overall stoichiometry is [\cvo{}]$_i$[\lao{}]$_{j-1}$CaAlO$_3$.
One can expect that the charge required to compensate the formal charge of the interface layer will come mainly from the partially filled $d$ states of the V cations, thus either depleting ($p$-type interface) or doping ($n$-type interface) the V $d$ states.
The symmetry of the unit cell enforces a vanishing electric field in the middle of both the CaVO$_3$ and LaAlO$_3$ layers, with a potential electric field emerging in the interfacial region, as indicated by the schematic electrostatic potential sketched in Fig.~\ref{fig:slabs}a).

The need to compensate the formal charges of the interface layers can be avoided in the alternative setup shown in Fig.~\ref{fig:slabs}b)~\cite{Lee/Demkov:2008}.
Here, a symmetric film of the $2+$/$4+$ compound in the center is sandwiched between an even number of positively and negatively charged sublayers of the $3+$/$3+$ material on both sides, which are further separated by a sufficient amount of vacuum ($>10$\,\AA).
This setup allows for the correct stoichiometry of the $3+$/$3+$ component, and thus all ions can in principle assume a bulklike valence without violating the overall charge neutrality. However, this comes at increased computational cost, due to the almost doubled size of the problem.
We note that even though in this case there is no apparent need for charge compensation at the interface, the system is nevertheless free to rearrange its charges to the most favorable configuration.
Due to the symmetry of the cell, the electric field is enforced to be zero in the middle of the $2+$/$4+$ component and the vacuum region, but can be non-zero in the polar component.

Finally, one can use an asymmetric setup, shown in Fig.~\ref{fig:slabs}c), which results from a simple stacking of $i$ full perosvkite units of CaVO$_3$ and $j$ full units of LaAlO$_3$, and contains both types of interfaces ($n$- and $p$-type).
From a stoichiometry perspective this setup in principle allows for a bulklike valence on all ions.
However, as already mentioned above, the charged sublayers in LaAlO$_3$ lead to a built-in polarity, which can result in a non-zero electric field inside the LaAlO$_3$ layer, and, due to the periodic boundary conditions, an opposing field inside the CaVO$_3$ layer.
In practice, this field can be compensated by electronic as well as structural reconstruction mechanisms~\cite{Nakagawa/Hwang/Muller:2006}, which can make this a rather complex scenario.

Note that all multilayer types described above have been used throughout the literature, e.g., for \lao{}/\sto{} multilayers~\cite{Pentcheva/Pickett:2010}.

\paragraph{Octahedral rotations}
In order to include the effects of the octahedral tilts that are present in both bulk \cvo{} and \lao{}, we construct all supercells such that they are in principle compatible with the corresponding tilt systems, using a pseudo-tetragonal unit cell with $\sqrt{2} \times \sqrt{2}$ in-plane lattice vectors.
The effect of lattice mismatch is included by fixing the in-plane lattice parameters to the pseudo-tetragonal in-plane bulk lattice parameters of the substrate. 
In the case of \lao{}, we neglect the rhombohedral strain of the bulk system for simplicity, always enforcing orthogonality between the three lattice vectors defining the supercell.
Independent of the number of layers, a glide plane parallel to the $c$-axis is preserved, such that each $B$O$_2$ layer only contains one inequivalent $B$ site.

In the following, we use the convention of referring to the rotations of the oxygen octahedra around the in-plane direction as ``tilts'', and to the rotations around the $c$ axis of the unit cell as ``rotations'', characterized by angles $\phi$ and $\theta$, respectively. The former is defined in terms of the out-of-plane V-O-V bond angle $\Phi=\pi-2\phi$, while the latter is defined in terms of the in-plane O-O-O bond angle  $\Theta=\pi/2-2\theta$ (see, e.g., Refs.~\cite{Rondinelli/Spaldin:2011,Beck/Ederer:2019}).
Note that we always take $\phi$ as positive, while $\theta$ can be either positive or negative, to indicate an in-phase (same sign) or out-of-phase (different sign) stacking of these rotations in subsequent planes.

To allow for an unconstrained relaxation of tilts across the multilayers, i.e., cases (a) and (c) in Fig.~\ref{fig:slabs}, the total number of perovskite layers perpendicular to $c$ has to be even, i.e., $i+j = 2k$ with $k\in \mathbb{N}$. 
Regarding the rotations around $c$, one has to consider the different Glazer tilt systems~\cite{Woodward:1997} of the bulk materials.
For the case of \cvo{} and \lao{} these are $a^-a^-c^+$ and $a^-a^-a^-$, respectively.
Thus, the stacking of the in-plane rotations differ in the two materials, i.e., in-phase in \cvo{} versus out-of-phase in \lao{}.
To be compatible with the latter, the number of AlO$_2$ sublayers has to be odd in the symmetric setup, case (a), independent of how the rotations couple across the symmetric interfaces. Thus, for the symmetric \cvo{}/\lao{} multilayers, we always choose both $i$ and $j$ to be odd.
Furthermore, as shown in Sec.~\ref{subsec:CL}, the coupling of the octahedral rotations across the interface in fact depends on the interface termination and differs for the $n$-type and $p$-type interface, such that in the asymmetric setup, case (c), the evolution of rotations is unrestricted for an even number of AlO$_2$ sublayers.
For case (b), any tilt pattern can establish naturally upon relaxation of the atomic coordinates due to the presence of the vacuum layer. However, choosing an even number of \cvo{} layers preserves a mirror plane in the central CaO layer.

\sto{} is cubic ($a^0a^0a^0$) at room temperature, but develops anti-phase rotations  ($a^0a^0c^-$) at low temperatures. Since DFT calculations correspond to the ground state at 0\,K, the corresponding rotations (or tilts) can also emerge in our \cvo{}/\sto{} multilayers.
To suppress this at least partially, we use an odd total number of perovskite layers with an even (odd) number of \cvo{} (\sto{}) layers. This also preserves a mirror plane within the central CaO layer.

In our calculations for \cvo{}/\sto{}, we always use a CaO/TiO$_2$-terminated interface, which corresponds to the most common surface termination of a \sto{} substrate observed experimentally~\cite{Kawasaki_et_al:1994}.
Analogously, we mostly focus on the CaO/AlO$_2$ ($p$-type) interface termination also for the case of \cvo{}/\lao{}, but we also discuss the VO$_2$/LaO-terminated ($n$-type) interface, to demonstrate potential differences resulting from different surface termination of the substrate.

\subsection{DFT+DMFT method}
\label{subsec:DMFT}

As already mentioned in the introduction, the dynamical correlation effects that underlie the formation of a Mott gap in the absence of any magnetic or orbital long range order need to be accounted for in a suitable way.
Here, we use a combination of DFT and DMFT, which has proven successful in describing the strain- and finite size-induced MIT in \cvo{}~\cite{Beck_et_al:2018}.

The general procedure, applied to all multilayers if not otherwise noted, is the following.
First, the cell parameter along $c$ and all internal coordinates are relaxed within DFT, while keeping the in-plane lattice parameters fixed.
We then identify a minimal correlated subspace consisting of the V-\ttg{}-dominated bands around the Fermi level., 
The corresponding Kohn-Sham states are converted to a local basis set of three \ttg{}-like maximally localized Wannier functions (MLWFs)~\cite{Marzari_et_al:2012} per V site.
Finally, the effects of the local Coloumb repulsion, $U$, and the Hund's interaction, $J$, within the correlated subspace are incorporated within subsequent DMFT calculations.
Depending on the specific multilayer and the values of the interaction parameters $U$ and $J$, this may open up a gap in the V \ttg{} states of \cvo{}, but we note that all pure DFT results are obtained for metallic \cvo{}.

For all DFT calculations we use the \textsc{Quantum~ESPRESSO} package~\cite{Giannozzi_et_al:2009}, the generalized gradient approximation according to Perdew, Burke, and Ernzerhof (PBE) for the exchange-correlation functional~\cite{Perdew/Burke/Ernzerhof:1996}, and scalar-relativistic ultrasoft pseudopotentials.
The 3s and 3p semicore states of Ca, V and Ti, the 4s and 4p semicore states of Sr, and the 5s and 5p semicore states of La, are included in the valence, while the empty La-4f states are not included.
The plane wave kinetic energy cutoff is set to 70\,Ry for the heterostructures which include \lao{}, and to 60\,Ry otherwise. A Monkhorst-Pack $6 \times 6 \times 4$ $k$-point grid is used for the smallest 20 atoms unit cells, and scaled accordingly in the $k_c$-component for the larger supercells. 
The Brillouin-zone integration is carried out using a Methfessel-Paxton smearing parameter of 0.01\,Ry.
Atomic positions are relaxed until all force components are smaller than 1 mRy/$a_0$ ($a_0$: Bohr radius). 
If the unit cell does not contain a vacuum layer, additionally the stress along $c$ is relaxed (while keeping the in-plane lattice parameters fixed), until the corresponding stress tensor component is smaller than 0.1\,kbar.
For the slab unit cells containing a vacuum layer, the \lao{} surfaces are separated by $>10$\AA{} in order to avoid any spurious interactions across the vacuum region.

The construction of maximally-localized Wannier functions for the Kohn-Sham bands defining the correlated subspace is performed with the \textsc{Wannier90} code~\cite{Mostofi_et_al:2008}.
In the case of \cvo{}/\lao{}, the relevant V-\ttg{}-dominated bands are well separated from V-\eg{}-dominated or \lao{}-related bands. 
\sto{}, on the other hand, exhibits low-lying empty Ti bands, that we include in the MLWF manifold in order to obtain consistent spreads of the MLWFs, which will be discussed further in Sec.~\ref{subsec:CS}.

The DMFT self-consistency cycle, relating the local blocks of the lattice Green's function 
to the corresponding effective impurity problems, is implemented using the TRIQS/DFTTools libraries~\cite{aichhorn_dfttools_2016}.
The effective impurity problems for the different symmetry-inequivalent sites are constructed from the Green's function based on the DFT Hamiltonian expressed in a suitable MLWF basis, supplemented with a local electron-electron interaction minus a double-counting correction.
To solve the impurity problems we use the continuous-time quantum Monte Carlo hybridization-expansion solver implemented in TRIQS/CTHYB~\cite{Seth2016274} at the inverse temperature, $\beta=1/(k_{\text{B}}T) = 40$ eV$^{-1}$.
The Hamiltonian describing the local electron-electron interaction is parameterized in the Slater-Kanamori form and includes both spin-flip and pair-hopping terms~\cite{Castellani/Natoli/Ranninger:1978,Vaugier/Jiang/Biermann:2012}. The Hund's coupling parameter is set to \mbox{$J=0.65$\,eV}, as used in other studies of early transiton metal oxides~\cite{Dang_et_al:2014,Beck_et_al:2018}, while the value of the Hubbard $U$ is varied systematically to study qualitative trends for the different multilayer geometries.
The orbital occupations are obtained from the imaginary time Green's function as \mbox{$n = -G(\beta)$}.
The spectral weight at the Fermi level, \mbox{$\bar{A}(0) = - \tfrac{\beta}{\pi} \textrm{Tr}[G(\beta/2)]$}
represents a measure of the metallic or insulating character of the system. 
Alternatively, the full frequency spectral function, $A(\omega)$, is calculated via analytic continuation using the maximum entropy method~\cite{jarrell1996bayesian}.
All results presented here are obtained from ``one-shot'' DFT+DMFT calculations, i.e., neglecting the effect of charge self-consistency.

%%%%%%%%%%%%%%%%%%%%%%%%%%%%%%%%%%%%%%%%%%%%%%%%%%%%%%%%
\section{Results and discussion}
\subsection{\cvo{}/\sto{}}
\label{subsec:CS}

\subsubsection{Structural properties and charge transfer -- DFT results}

As mentioned in Sec.~\ref{sec:cm}, the absence of a polar discontinuity in \cvo{}/\sto{} heterostructures does not lead to any complications related to the specific multilayer geometry. Therefore, we choose to use the computationally most efficient symmetric multilayer with two chemically identical interfaces (corresponding to case (a) in Fig.~\ref{fig:slabs}).
Our calculated pseudo-tetragonal lattice parameters for bulk \cvo{} and \sto{} are \mbox{$\atet=3.784$\,\AA} and \mbox{$\atet=3.938$\,\AA}, respectively, which lead to a relatively large lattice mismatch corresponding to a tensile strain of $4.1\%$ in \cvo{} when fixing the in-plane cell parameters to that of \sto{}.
This is slightly larger than for the corresponding experimental lattice parameters, \mbox{$\atet=3.771$\,\AA} and \mbox{$\atet=3.905$\,\AA}~\cite{Falcon_et_al:2004}, which would result in a tensile strain of $3.6$\,\%.
The difference is due to different systematic errors in the calculated lattice parameters for the two materials, which are in the typical range of deviations for the PBE functional.
Here, \mbox{$\atet{}=(a+b)/2\sqrt{2}$} refers to the pseudotetragonal cell parameter calculated from the average of the lattice parameters $a$ and $b$.

\begin{figure}
    \centering
    \includegraphics[width=\linewidth]{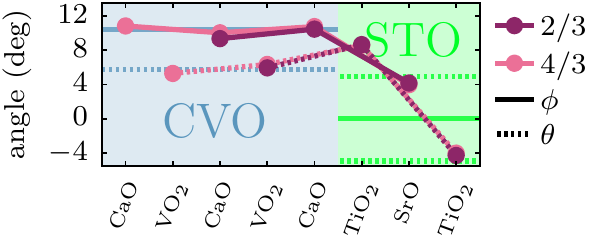}
    \caption{%
    Evolution of octahedral tilt ($\phi$) and rotation ($\theta$) angles along the $c$-direction in the \cvo{}/\sto{} multilayers, represented by solid and dashed lines, respectively.
    According to their definition (see Sec.~\ref{subsec:DFT}), tilt angles are assigned to the $A$O layers and rotation angles to the $B$O$_2$ layers.
    The corresponding values for the bulk materials are shown as blue (strained \cvo{}) and green (\sto{}) horizontal lines.
    Note that only half of each layer is shown, the other half is related by mirror symmetry.
    }
    \label{fig:ang_CS}
\end{figure}

We first analyze the octahedral tilts and rotations of the relaxed structures, using the angles defined in Sec.~\ref{subsec:DFT}.
Manipulating the degree of these octahedral tilts and rotations through interfacial octahedral coupling has recently been suggested as a possible design tool for exploring new emerging phases in oxide heterostructures~\cite{Rondinelli/May/Freeland:2012}.
In Fig.~\ref{fig:ang_CS} we plot the layer-dependent tilt and rotation angles in \cvo{}/\sto{} multilayers for two different layer thicknesses.

In spite of the fact that we are using an odd total number of perosvkite units along $c$, which prohibits the development of continuous tilts throughout the periodic multilayer, the octahedral rotations are surprisingly bulklike after the relaxation, both in the case of two and four layers of \cvo{}.
In particular, both $\theta$ and $\phi$ in the \cvo{} part are nearly unaffected by the presence of the \sto{} ``substrate'' and the resulting change in tilt pattern and the mismatch of the corresponding angles across the interface.
To better understand the role of the octahedral rotations in the \sto{} layer, we also perform relaxations (not shown here) of a multilayer in which the atomic positions corresponding to \sto{} are fixed according to a locally tetragonal structure without tilts/rotations, apart from the interfacial O atoms in the TiO$_2$ layers.
This results in a reduction (increase) of up to $14\%$ ($9\%$) of the tilts (rotations) in \cvo{}, seemingly shifting the degree of octahedral distortions from the tilts towards the rotations, as these are only weakly coupled between neighboring layers.
However, these structural changes do not significantly affect the results of the subsequent DMFT calculations. We therefore do not expect a strong influence of the octahedral rotations inside the \sto{}, which are absent in experiments conducted at room temperature, on the electronic properties of the \cvo{}/\sto{} interface.

\begin{figure}
    \centering
    \includegraphics[width=\linewidth]{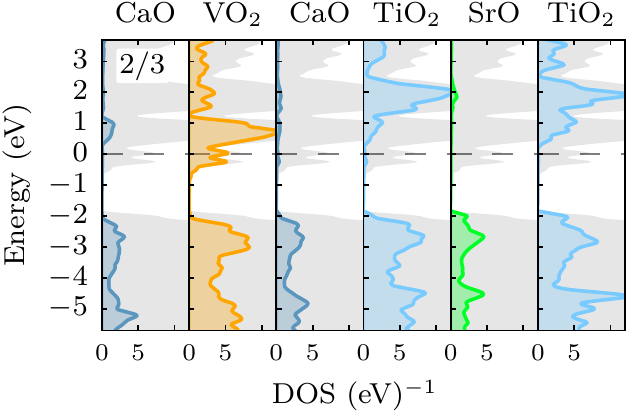}
    \caption{%
    Layer-resolved density of states (DOS) in the $2/3$ \cvo{}/\sto{} multilayer with the Fermi level at zero energy.
    The grey-shaded area in each panel indicates the total DOS of the slab.
    The correlated subspace is constructed within an energy window of $[-1,2.5]$\,eV, which includes the partially filled V-\ttg{} dominated bands and the (empty) Ti-\ttg{} dominated bands.
    Note that only half of each layer is shown, the other half is related by mirror symmetry.
    }
    \label{fig:dos_CS23}
\end{figure}

As noted before, the empty Ti \ttg{} states are relatively low in energy, giving rise to the general possibility of charge spilling from the V-$d$ valence states into the Ti-\ttg{} states across the interface within the heterostructures.
However, we find that the band alignment is such that this charge transfer into the \sto{} layers is very small ($<0.05$ electrons in the interface layer, zero in the sub-interface layer), as shown in Fig.~\ref{fig:dos_CS23}, such that we neglect it in the following.
Nevertheless, if we construct MLWFs only for the V-\ttg{} states (in the energy range between $-1$ and about 1.5\,eV in Fig.~\ref{fig:dos_CS23}), then the Wannier functions located on the V sites close to the interface exhibit pronounced ``tails'' on the neighboring Ti sites, which are due to hybridization with the Ti states across the interface.
These tails significantly increase the quadratic spread of the corresponding Wannier functions.
Thus, in order to have consistent spreads and thus results that are directly comparable to bulk CaVO$_3$ and CaVO$_3$/\lao{} heterostructures, we also include the Ti-\ttg{} bands in the construction of the Wannier basis, using an energy window between $-1$ and 2.5\,eV, but then treat the Ti states as ``uncorrelated'', i.e., they are included in the lattice Green's function but are considered as non-interacting.

\subsubsection{Metal-insulator transition -- DMFT results}

\begin{figure}
    \centering
    \includegraphics[width=\linewidth]{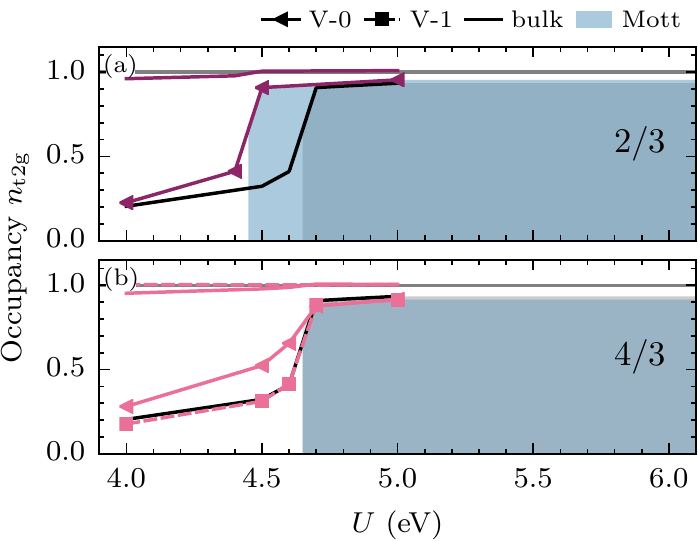}
    \caption{%
    DMFT results for the 2/3 (top) and 4/3 (bottom) \cvo{}/\sto{} multilayers obtained for different values of the interaction parameter $U$.
    Lines with markers show the orbital polarization, while lines without markers depict the total site occupation of the respective layers.
    Different layers are denoted as V-$i$, where $i$ refers to the proximity to the interface.
    The  orbital polarization of strained bulk is shown as black solid line.
    The Mott insulating regime for the different layers (strained bulk) is indicated by the blue-shaded (grey) area.
    }
    \label{fig:op_CS}
\end{figure}

The results of our DMFT calculations for the \cvo{}/\sto{} multilayers are summarized in Fig.~\ref{fig:op_CS}, where the (layer-dependent) total occupations of the three V-\ttg{} orbitals (colored lines without markers) are plotted as function of $U$, together with the corresponding orbital polarization (colored lines with markers), defined as the difference between the highest and lowest occupation of the three individual orbitals. The latter is also compared to the strained bulk case (solid black line). In addition, the insulating range is indicated as shaded areas.

It can be seen that strained bulk becomes insulating for $U \geq 4.7$\,eV and that the metal-insulator transition is accompanied by a strong enhancement of the orbital polarization, with one orbital essentially completely filled and the other two empty, consistent with previous work~\cite{Beck_et_al:2018}. 
Note that often a value of $U \approx 5$\,eV has been suggested as realistic value for \cvo{} and related systems~\cite{Pavarini_et_al:2004}, so that bulk \cvo{} would be insulating for a strain of $\sim 4$\,\%.

As shown in Fig.~\ref{fig:op_CS}a for the 2/3 multilayer containing only two VO$_2$ layers, \cvo{} in this case becomes insulating at $\umit{}=4.5$ eV,
which is only 0.2\,eV lower than for strained bulk, indicating that the effect of the confinement due to the finite thickness of the \cvo{} layer is rather weak.
This can be understood from the fact that the two orbitals that respond to the finite size, i.e. $d_{xz}$ and $d_{yz}$ orbitals, are unoccupied in the insulating state, while the completely filled $d_{xy}$ orbital is only affected by the strain, and not by the thickness of the film.
One can also observe that, as already mentioned above, the charge spilling into the \sto{} is indeed negligible, indicated by the total occupation of approximately 1 in the (two symmetry-equivalent) VO$_2$ layers, which becomes exactly 1 in the insulating regime.

Finally, in the 4/3 multilayer, i.e., with only four layers of \cvo{}, the system behaves completely bulklike again, albeit with a slightly enhanced orbital polarization in the interface layer in the metallic regime, as demonstrated in Fig.~\ref{fig:op_CS}b.
One can thus conclude that the finite size effect vanishes already for four layers within the multilayers, and that since the MIT in the unstrained bulk material occurs only for $U \geq 5.6$\,eV~\cite{Beck_et_al:2018} the largest effect on the MIT is due to the epitaxial strain.

In comparison, the MIT for a free-standing slab containing two monolayers of (unstrained) \cvo{} was found to be shifted towards lower $U$ values by more than 1\,eV compared to (unstrained) bulk~\cite{Beck_et_al:2018}. This huge effect has been attributed mainly to a strong enhancement of the ``crystal-field'' splitting, i.e., the difference in the orbital energies between the three \ttg{} levels due to the symmetry reduction in the free-standing slabs. For thicker free-standing slabs, the enhancement of the crystal-field splitting was only significant in the surface layers. In the present multilayer systems, an enhancement of the crystal-field splitting in the surface, respectively interface VO$_2$ layer, is essentially absent, due to the presence of \sto{}, which, in a very simple point charge picture with nominal valences, would be indistinguishable from \cvo{}.

Thus, we conclude that, embedded within an isovalent \sto{} ``substrate'', \cvo{} essentially does not exhibit any noticeable finite size effects, even down to a thickness of only 4 perovskite units.
Here, the tensile strain is the dominant effect, leading to an enhanced orbital polarization in favor of a half-filled in-plane $d_{xy}$ orbital, which is already present within the strained bulk system.
The results show that the presence of the \sto{} substrate significantly weakens the finite thickness effects obtained for ultrathin free-standing films of \cvo{}. 
For a more realistic thin film geometry with \sto{} substrate on one side and vacuum on the other side of the \cvo{} layer, we expect a behavior intermediate between the free-standing slab and the multilayer.
This supports previous conclusions from Ref.~\onlinecite{Beck_et_al:2018} that the thickness-dependent metal-insulator transition observed in thin films of \cvo{} grown on \sto{}~\cite{Gu_et_al:2013} is mainly due to substrate-induced strain and less related to finite size effects.
However, we note that for unstrained or compressively strained \cvo{}, i.e for cases without a preferential $d_{xy}$ occupation, stronger finite size effects may be expected.

\subsection{\cvo{}/\lao{}}
\label{subsec:CL}

\subsubsection{Structural properties and interfacial doping -- DFT results}

\paragraph{Structural properties}
We first discuss the implications of the supercell geometry on the lattice structure.
The pseudo-tetragonal cell parameter calculated for bulk \lao{} amounts to \mbox{$\atet=3.813$}\,\AA, which agrees within 0.6\,\% with the experimental value of $\atet=3.789$\,\AA~\cite{Ohtomo/Hwang:2004}.
If we thus fix the lateral dimensions of the supercells to the lattice parameter of the substrate, the \cvo{} films are coherently strained by $0.8\%$. 

\begin{figure}
    \centering
    \includegraphics[width=\linewidth]{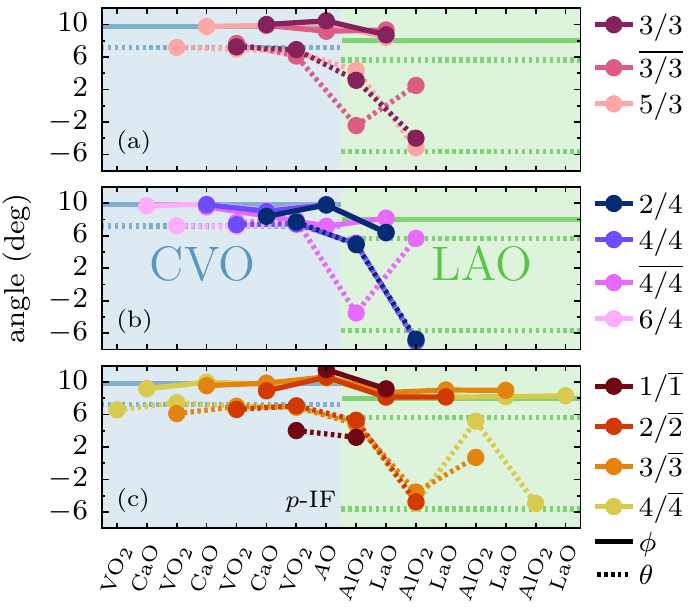}
    \caption{%
    Evolution of octahedral tilt and rotation angles, $\phi$ (solid lines) and $\theta$ (dashed lines), respectively, in the different \cvo{}/\lao{} multilayers along the $c$-direction.
    The horizontal lines indicate the corresponding bulk values for strained \cvo{} (blue), and \lao{} (green).
    (a) Symmetric multilayers.
    (b) Symmetric, vacuum-separated slabs.
    (c) Asymmetric mutlilayers.
    }
    \label{fig:ang}
\end{figure}

In Fig.~\ref{fig:ang} we show the evolution of the tilt ($a^-a^-c^0$) and rotation ($a^0a^0c^{\pm}$) angles, $\phi$ and $\theta$, respectively, for the various \cvo{}/\lao{} multilayers after full relaxation of all atomic coordinates and the $c$-component of the unit cell, as described in Sec.~\ref{subsec:DFT}.
Here, we use the notation $i/j$ for the symmetric multilayers with $p$-type interfaces, analogous to the case of \cvo{}/\sto{} described in the previous section, whereas we denote the corresponding cases with $n$-type interfaces as $\overline{i/j}$, and the asymmetric multilayers as $i/\overline{j}$, to indicate that the latter always contain one $p$- and one $n$-type interface.
Overall, we find in all three cases, (a)-(c), i.e. the three different multilayer geometries introduced in Sec.~\ref{sec:cm} and Fig.~\ref{fig:slabs}, that the tilts and rotations in \cvo{} are already very similar to their bulk values at the sub-interface layer, i.e. one layer away from the interface.
The only exception is perhaps the (somewhat extreme) $1/\overline{1}$ case, where the tilts are increased with a simultaneous reduction of the rotation angles.
For the vacuum slabs in Fig.~\ref{fig:ang}b), the effect of the \lao{} surface shows up only minimally in the \lao{} layers.
We therefore conclude that the different multilayer geometries have no significant effect on the octahedral tilts and rotations.

An interesting aspect emerging from these calculations is the coupling of the in-plane rotations across the interface.
In bulk \cvo{}, the rotations in subsequent layers couple in-phase, while in bulk \lao{} they couple out-of-phase.
From our calculations, we consistently find that, if the multilayer geometry allows, the coupling of the rotations across the interface is in-phase for the $p$-type and out-of-phase for the $n$-type interface.
Thus, if the interfacial $A$O layer is LaO, then the rotations couple as in \lao{}, whereas if it is CaO, then they couple as in \cvo{}.
This means that in the asymmetric case, where both types of interfaces are present, the coupling depends on the number of \lao{} layers.
For an even number of \lao{} layers, i.e. $2/\overline{2}$ and $4/\overline{4}$, the rotation of the two interfacial Al octahedra are opposite, such that the periodicity allows the aforementioned in-phase coupling at the $p$-type and anti-phase coupling at the $n$-type interface, respectively.
On the other hand, in both the $1/\overline{1}$ and $3/\overline{3}$ superlattices the interfacial Al octahedra would have the same sense of rotation, which does not allow for one in- and one anti-phase rotation coupling, resulting in a state with completely suppressed rotations at the $n$-type interface.
These findings, however, seem to affect only \lao{}, while the octahedral distortions in \cvo{} are only marginally altered.
We note that we also calculated symmetric slabs with an even number of \lao{} layers, or an odd total number of layers to observe potential structural differences.
While this can suppress or enhance tilts and rotations noticeably in \lao{}, we found that, again, there is no strong influence on the \cvo{} film.
We further note that these results are independent of how octahedral rotations are initialized in the calculations.

\paragraph{Occupations and crystal-field splitting}
\label{subsec:lao-occ}

Due to the substantially larger band gap of \lao{} compared to \sto{}, the V \ttg{} bands that define the correlated subspace for the DMFT calculations are completely separated from other bands at higher and lower energies in the \cvo{}/\lao{} multilayers.
However, the polar discontinuity at the interface can induce changes in the local occupations of the V-\ttg{} Wannier functions and the corresponding crystal-field splitting, as obtained from our DFT calculations.

\begin{figure}
    \centering
    \includegraphics[width=\linewidth]{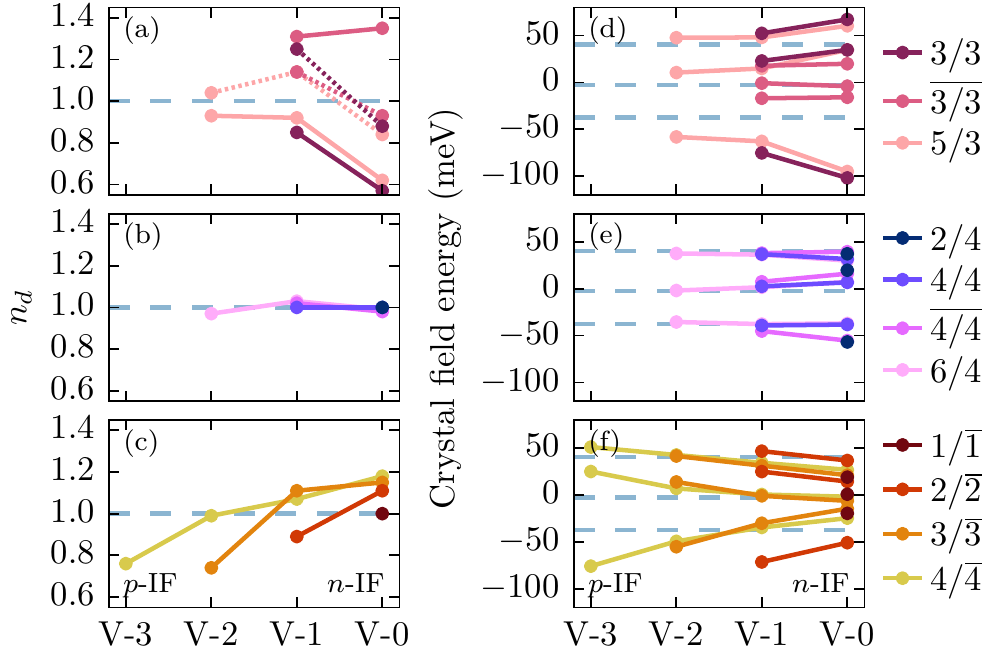}
    \caption{%
    Layer-dependent site occupations (left) and crystal-field splittings (right) of the V-\ttg{} MLWFs as a function of distance from the interface.
    Results for the symmetric multilayers are shown in panels (a) and (d), for the vacuum slabs in panels (b) and (e), and for the asymmetric multilayers in (c) and (f).
    The corresponding values for strained bulk \cvo{} are indicated as dashed horizontal blue lines.
    The dotted lines in panel (a) correspond to calculations with artificially adjusted filling of the V-\ttg{} bands (see text).}
    \label{fig:dft_set}
\end{figure}
In Fig.~\ref{fig:dft_set}a)-c) we plot the total occupation of the V-\ttg{}-manifold in each layer. In bulk \cvo{} this occupation is exactly equal to 1, as indicated by the blue dashed lines.
It can be seen that in the symmetric multilayers with a $p$-type interface (cases $3/3$ and $5/3$ in Fig.~\ref{fig:dft_set}a)), the V in the VO$_2$ layer closest to the interface is depleted by $\sim 0.4$ electrons, while the occupation of the V sites further away from the interface are rather close to the bulk occupation.
This is consistent with the expectations discussed in Sec.~\ref{subsec:DFT}, i.e., that in order to compensate the formal charge of the additional AlO$_2$ layer, the V-\ttg{} bands are depleted in total by one electron per V ($1/2$ an electron per V on each interface). The hole-doping seems to be confined to the interface layer, with no significant difference between the $3/3$ and the $5/3$ cases, i.e., for three or five VO$_2$ layers.
Comparing this with the $n$-type interface (labeled $\overline{3/3}$), one recognizes that in this case, as expected, the additional LaO layer results in an excess electron within the V-\ttg{} states. However, this additional electron appears to be more evenly spread over the different VO$_2$ layers compared to the hole distribution for the $p$ type interface.

To better understand the effect of the excess electron or hole at the $n$- and $p$-type interfaces, respectively, resulting from the overall stoichiometry in the symmetric multilayers, we artificially add or remove electrons from the MLWFs, to obtain a bulklike filling of the V $d$ bands.
The resulting occupations are shown as dotted lines in Fig.~\ref{fig:dft_set}a). 
There is still a noticeable variation in the charge distribution across the different layers, with a weak depletion of the interface layer for both types of interfaces, resulting in excess charge in the subinterface layer.

The crystal-field energies of the MLWFs for the symmetric multilayers, calculated as the diagonalized on-site energies, are shown in Fig.~\ref{fig:dft_set}d), where the crystal-field energies are centered around their mean energy per layer, as a function of the layer depth.
Compared to the strained bulk reference (horizontal dashed lines), the crystal-field splitting is strongly enhanced at the $p$-type interface, with one orbital being noticeably lower in energy than the other two. This generally favors a strong orbital polarization and thus the development of a Mott-insulating state~\cite{Sclauzero/Dymkowski/Ederer:2016}.
In the $n$-type slab ($\overline{3/3}$), on the other hand, the crystal-field splitting is strongly reduced, suggesting a reduced orbital polarization and a more metallic state (see Sec.~\ref{subsec:lao-dmft}).

The electron- or hole-doping of the V states observed in the symmetric multilayers is completely absent in the symmetric vacuum slab, shown in Fig.~\ref{fig:dft_set}b). Here, the occupations are completely bulklike (equal to one) for both the $p$-type and the $n$-type interface, and for all layer depths and slab sizes. As outlined in Sec.~\ref{subsec:DFT}, due to the overall stoichiometry of these slabs, all ions can adopt the same formal valences as in the corresponding bulk materials, and, as seen in Fig.~\ref{fig:dft_set}b), no charge transfer occurs at the interface.
This is also consistent with the corresponding crystal-field energies shown in Fig.~\ref{fig:dft_set}e), which are essentially unchanged compared to bulk values already one layer away from the interface, and only slightly affected in the interface layer, independent of the interface type.

Finally, for the asymmetric multilayers shown in Fig.~\ref{fig:dft_set}c), we find a depletion of the V-\ttg{} states at the $p$-type interface and an excess occupation at the $n$-type interface, consistent with the internal potential gradient of this cell-type.
Notably, it appears that the hole-doping at the $p$-type interface is limited to a single VO$_2$ layer, while the corresponding excess electronic charge is spread out over at least two layers close to the $n$-type interface, similar to what is observed for the symmetric multilayers in Fig.~\ref{fig:dft_set}a).
The corresponding crystal-field energies are shown in Fig.~\ref{fig:dft_set}f).
Here, the internal gradient seems to induce a monotonous increase of the crystal-field splitting, starting from a reduced splitting compared to bulk at the $n$-type interface towards an enhanced splitting at the $p$-type interface.
This enhancement/reduction is also in agreement with the observed changes in the symmetric multilayers (see Fig.~\ref{fig:dft_set}d)).

Overall we find essentially  bulklike occupations in the symmetric vacuum-separated slabs, while the electronic properties in the multilayer geometries differ significantly from the bulk material, and are also strongly dependent on the specific interface termination.
In other words, exchanging the interfacial $A$-site atom from Ca to La has a strong effect on both the occupancy and the crystal-field energies of the V-sites in the multilayer cases.
Since the octahedral tilts and rotations are not strongly affected by the presence of the interface, and furthermore behave very similar for the different multilayer geometries, they can be ruled out as source for the interfacial electronic reconstruction.
Nevertheless, we now briefly discuss alternative lattice reconstruction mechanisms, that may provide further insights regarding the modifications of the electronic structure.

First, we note that in all multilayer calculations, the V atoms off-center relative to their surrounding oxygen octahedron away from (towards) the $n$-type ($p$-type) interface, which also compensates, at least partially, the corresponding polar discontinuity.
Second, the magnitude of this off-centering is stronger at the $n$- than at the $p$-type interface.
This is in contrast to \lao{}/\sto{}, where structural reconstruction is dominant at the $p$-type interface, since hole-doping is not possible for Ti$^{4+}$. 
On the other hand, the creation of holes in V$^{4+}$ appears very effective, since the doping at the $p$-type interface is essentially localized within a single interfacial layer, compared to the more delocalized electron doping at the $n$-type interface.
Third, the average out-of-plane O-V-O distance is enhanced (reduced) at the $n$-type ($p$-type) compared to the bulk value.
It is conceivable that a reduction of the corresponding V-O bond length distance, which coincides with a lower valence of the intermediate V cation, is less favorable than an increased bond length, resulting in a minimisation of the number of sites with lower than nominal valence.

Altogether, taking into account both the off-centering and the bond length variation, these structural changes could contribute to the interfacial reconstruction, in addition to the reported trends in the electronic structure.
While these effects are likely related to the non-polar/polar nature of the interface, interestingly, they are not present in the vacuum slab, indicating again the important role of the specific electrostatic boundary conditions.

\subsubsection{Metal-insulator transition -- DMFT results}
\label{subsec:lao-dmft}

Next, we present the results of our DFT+DMFT calculations for \cvo/\lao{} heterostructures and discuss how the supercell geometry affects the tendency of the \cvo{} films to form a Mott-insulating state.

\paragraph{Symmetric multilayer}

\begin{figure}
    \centering
    \includegraphics[width=\linewidth]{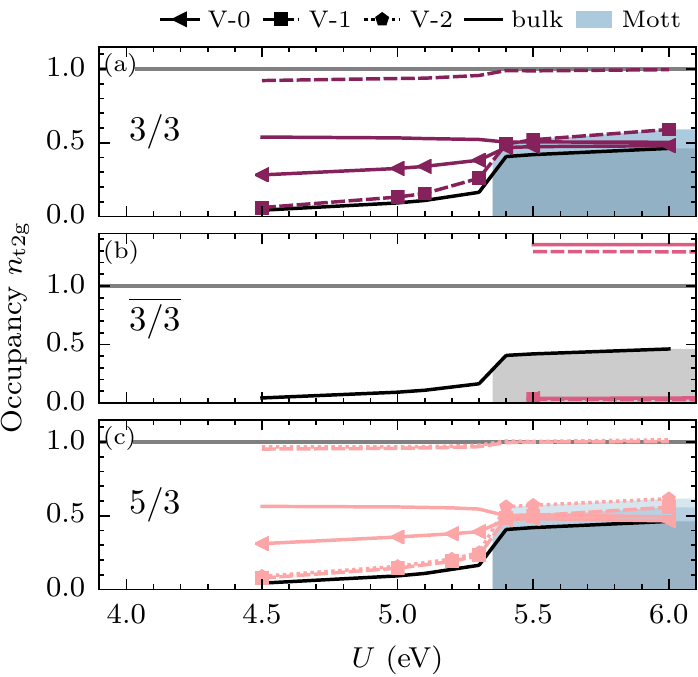}
    \caption{%
    DFT+DMFT results for the symmetric \cvo{}/\lao{} multilayers, obtained as function of the Hubbard parameter $U$.
    The lines with markers show the orbital polarization, while the lines without markers depict the total site occupation of the different layers, denoted as V-$i$, where $i$ refers to the proximity to the interface.
    The orbital polarization of the strained bulk system is shown as black solid line.
    The  Mott insulating regime for the different layers (strained bulk) is represented by the blue (grey) shaded area.
    }
    \label{fig:op_sym_dft}
\end{figure}

As discussed in the previous section, the total occupation of the V states in the symmetric multilayer is off by half an  electron or hole per V per interface.
Since the Mott-insulating state requires an integer filling per site (in this case one electron per Vanadium site), it is a priori impossible to achieve this for all sites with symmetric interfaces. This is confirmed by our calculations, where, depending on the strength of the interaction parameter $U$, an insulating state is only achieved for the inner VO$_2$ layers, whereas the V sites in the interface layers remain metallic for any $U$, due to their non-integer occupation.

Fig.~\ref{fig:op_sym_dft} shows the orbital polarization, i.e., the occupation difference between the most and least occupied \ttg{} orbitals per site, as well as the total site occupation, in comparison to the orbital polarization of strained bulk \cvo{}.
Due to the lower strain compared to the cases with the \sto{} substrate discussed in Sec.~\ref{subsec:CS} (0.8\,\% versus 4.1\,\%) the bulk system becomes insulating at a higher $U_\text{MIT}$ of 5.4\,eV, and the orbital polarization in the insulating state is reduced to about 0.5.
For both multilayers with $p$-type interfaces, 3/3 and 5/3, one can see that, as expected, the inner layers (V-1 and V-2) adopt the exact nominal occupancy of 1 when undergoing the metal-insulator transition, whereas the interface layer (V-0) is filled with 0.5 electrons and keeps a finite spectral weight at the Fermi level.
Within the accuracy of our calculations, both the 3/3 and the 5/3 multilayers exhibit the same \umit{} as the strained bulk system. 

Strikingly, changing the interface termination to $n$-type (case $\overline{3/3}$, see Fig.~\ref{fig:op_sym_dft}b) results in very different behavior. Here, also the inner layer stays metallic over the full range of $U$ values. 
This is likely due to the strongly reduced crystal-field splitting (see Fig.~\ref{fig:dft_set}d), such that even a rather strong onsite Coulomb repulsion of up to 7.5\,eV is unable to generate an integer occupation of the middle layer (V-1).

We note that as mentioned in Sec.~\ref{subsec:lao-occ} we also performed DMFT calculations where we added an additional electron to the system after the construction of the MLWFs, to allow all V cations to adopt their bulklike stoichiometric nominal occupancies.
In the corresponding DMFT calculations, the systems with $p$-type interfaces exhibit a much stronger tendency to become insulating (\umit{} is shifted by $\approx0.5$ eV), since integer filling can now be achieved for all layers. In addition, the enhanced crystal-field splitting at the $p$-type interface results in a very strong orbital polarization.
On the other hand, analogous calculations for the $n$-type multilayer, with one electron removed from the system, show insulating behaviour for $U \geq 5.5$ eV, but again without orbital polarization.

While the calculations with adjusted electron count are somewhat artificial, these results nevertheless indicate that the charge-doping effects at both $n$- and $p$-type interfaces strongly favor the metallic state due to the resulting half-integer occupation of the V bands.
On the other hand, the finite size and strain effects are in principle expected to favor the insulating state.
Thus, there are two opposing effects at play here, which seem to essentially compensate each other for the $p$-type multilayers, resulting in the same \umit{} as the strained bulk system.
This is also supported by the following analysis of the symmetric, vacuum-separated slabs, where we already illustrated in Fig.~\ref{fig:ang}b) and Fig.~\ref{fig:dft_set}b) and e) that the effects of the interface are nearly negligible, and we will further demonstrate in the following that for this setup the influence of the interface is reduced to a mere finite size effect.

\paragraph{Symmetric vacuum-separated slabs}

\begin{figure}
    \centering
    \includegraphics[width=\linewidth]{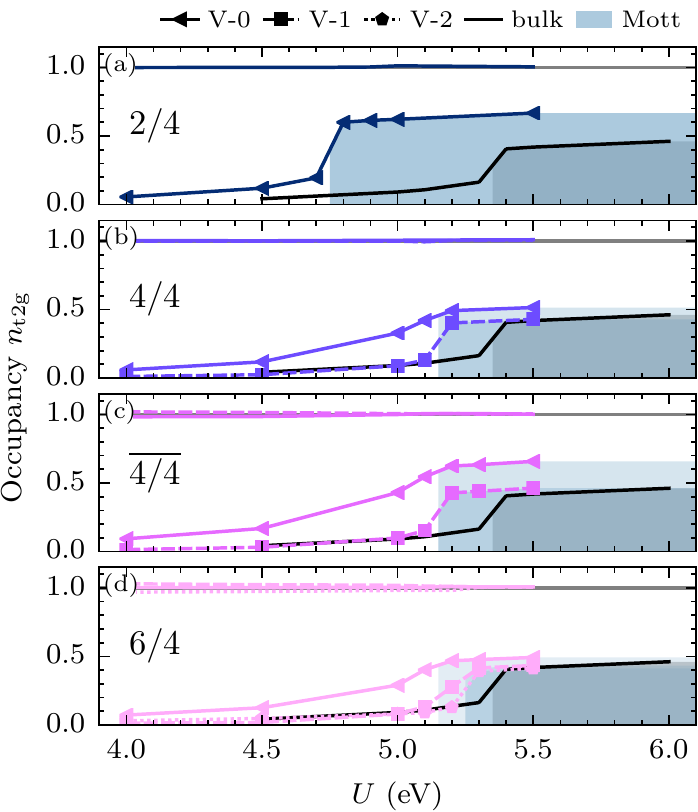}
    \caption{%
    DMFT results for the symmetric \cvo{}/\lao{} vacuum slabs.
    The lines with markers show the orbital polarization, while the lines without markers depict the total site occupation of the respective layers.
    The layers are indicated by V-$i$, where $i$ refers to the proximity to the interface.
    The strained bulk orbital polarization is shown as black solid line.
    The Mott insulating regime for the different layers (strained bulk) is represented as a blue-shaded (grey) area.
    }
    \label{fig:op_stoi}
\end{figure}

Our DFT+DMFT results for the symmetric vacuum-separated slabs are shown in Fig.~\ref{fig:op_stoi}.
For the thinnest \cvo{} film (case 2/4), the critical $U$ for the metal-insulator transition is significantly reduced compared to bulk to $\umit{}=4.8$\,eV. 
Upon increasing the \cvo{} thickness, this value steadily converges back to the corresponding bulk value.
Similarly, the orbital polarization is enhanced in the ultra-thin film compared to the bulk system, in line with a slightly enhanced crystal-field splitting shown in Fig.~\ref{fig:dft_set}e).
In contrast to the case of the symmetric multilayers, and consistent with the DFT occupations shown on Fig.~\ref{fig:op_sym_dft}, the results are independent of the interface termination, which can be seen by comparing the 4/4 slab ($p$-type interface, Fig.~\ref{fig:op_stoi}b) with the $\overline{4/4}$ case ($n$-type interface, Fig.~\ref{fig:op_stoi}c).
Even though the latter exhibits a slightly larger orbital polarization (due to the slightly larger crystal-field splitting), this has no noticeable effect on \umit{}.

It was shown in Ref.~\onlinecite{Beck_et_al:2018} that free-standing ultra-thin films of \cvo{} are insulating due to a combination of dimensional confinement as well as surface effects, resulting in an enhanced crystal-field splitting within the surface VO$_2$ layer. 
Compared to the free-standing case, the splitting in the outermost VO$_2$ layer is significantly reduced by the presence of the \lao{} substrate in the vacuum-separated slabs, and only slightly enhanced compared to bulk, such that the corresponding effect on the MIT is expected to be rather small.
Thus, with nominal occupations and bulklike crystal-field splittings, the difference between these slabs and the corresponding bulk system is essentially only the finite number of layers, free from any influence of the polar nature of the \lao{} substrate.

\paragraph{Asymmetric multilayers}

\begin{figure}
    \centering
    \includegraphics[width=\linewidth]{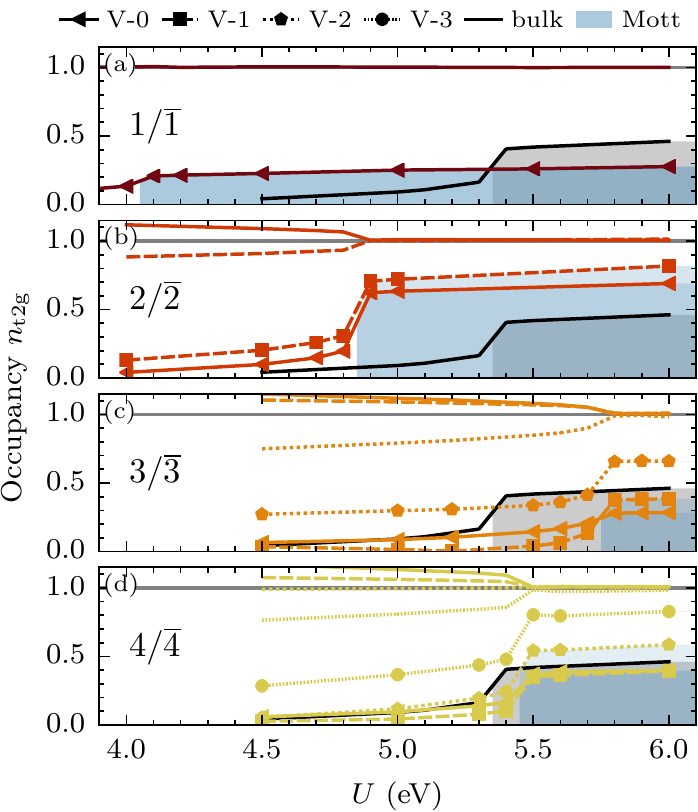}
    \caption{%
    DFT+DMFT results for the asymmetric \cvo{}/\lao{} multilayers.
    The lines with markers show the orbital polarization, while the lines without markers depict the total site occupation of the respective layers.
    The layers are indicated by V-$i$, where in this case $i$ refers to the proximity to the $n$-type interface.
    The strained bulk orbital polarization is shown as a black solid line.
    The Mott insulating regime for the different layers (strained bulk) is represented as a blue-shaded (grey) area.
    }
    \label{fig:op_asym}
\end{figure}
Finally, the asymmetric case, with both types of interfaces simultaneously present, is shown in Fig.~\ref{fig:op_asym}. 
Due to the redistribution of electrons from the $p$-type to the $n$-type interface, resulting from the polarity of the slab, several V sites exhibit non-integer occupation for small $U$, which can be expected to strongly favor a metallic state.
The extreme case of the $1/\overline{1}$ multilayer (see Fig.~\ref{fig:op_asym}a)) is an exception to this.
Here, the occupation of the single V layer remains fixed to 1, resulting in a strong shift towards the insulating state compared to strained bulk, with $\umit{}=4.1$\,eV.
This is due to the strongly reduced dimensionality for a single VO$_2$ layer, despite the reduced crystal-field splitting and the resulting small orbital polarization.
The finite size effect also seems to outweigh the effect of the charge redistribution in the $2/\overline{2}$ multilayer (see Fig.~\ref{fig:op_asym}b), since this case still has a \umit{} lower than the strained bulk system, albeit increased compared to the $1/\overline{1}$ case due to the larger film thickness.
Upon increasing $U$, the charge imbalance between the two V layers is gradually reduced and both layers become insulating with an occupation equal to 1 above \umit{}.

For the thicker multilayers, a qualitative change occurs from favoring the insulating state to favoring the metallic state compared to bulk.
This non-monotonous trend indicates the opposing effects of finite size and charge redistribution.
As shown in Fig.~\ref{fig:op_asym}c), the critical $U$ for three V layers is $\umit{}=5.8$\,eV, i.e. higher than in the bulk, while for four layers \umit{} is only 0.1\,eV larger than for bulk.
Thereby, the sites with lower occupancy in the metallic state and enhanced crystal-field splitting, i.e. those at the $p$-type interface, show a high orbital polarization (see Fig.~\ref{fig:dft_set}c) and f)), while the sites with higher occupation in the metallic state and reduced crystal-field splitting have a slightly reduced orbital polarization compared to bulk.

\begin{figure}
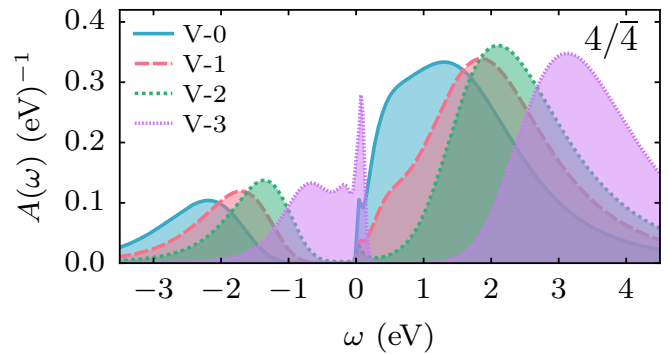

    \centering
    \includegraphics[width=\linewidth]{{{./sfunc_CL44}}}
    \caption{%
    Layer-resolved orbitally averaged spectral functions $A(\omega)$ for the asymmetric $4/\overline{4}$ multilayer at $U = 5.5$\,eV.
    The Mott gap within each layer is clearly visible, however, it is exceeded by the potential shift across the layers.
    }
    \label{fig:spec}
\end{figure}

Notably, all sites within the multilayer undergo the metal-to-insulator transition simultaneously at a certain value for the on-site Coulomb repulsion, even in the $4/\overline{4}$ case where the occupation of the V-2 layer is essentially equal to the nominal occupation already for small $U$.
However, even above \umit{}, the spectral weight at the Fermi level remains finite at the site that is located at the $p$-type interface (thus, no corresponding shading in the graphs), for both the $3/\overline{3}$ and $4/\overline{4}$ multilayers. 
While the local spectral function for each site exhibits a Mott gap, the total spectral function appears metallic. This is due to a continuous shift in energy of the local spectral densities caused by the potential gradient across the \cvo{} film, which is larger than the Mott gap. Consequently, the lower Hubbard band at the $p$-type interface (V-3) overlaps with the upper Hubbard band at the $n$-type interface (V-0), as shown in Fig.~\ref{fig:spec}.
Since the Fermi level is pinned to the bottom of the latter, the $p$-type interface thus remains metallic. 

%%%%%%%%%%%%%%%%%%%%%%%%%%
\section{Summary and Conclusions}
\label{sec:sum}

In this work, we have investigated effects resulting from the substrate-film interface on the structural and electronic properties of \cvo{} films for two different commonly used substrate materials, \sto{} and \lao{}. We have shown that the effect of these two substrate materials are fundamentally different, not only due to the strongly different lattice mismatch they impose on \cvo{}, but mostly due to the different polarity of the corresponding interfaces.

For the case of the \sto{} substrate, which, as \cvo{}, can be viewed as a stacking of charge-neutral layers, there is no polar discontinuity at the interface, and no charge transfer occurs between the partially filled V-\ttg{} bands of the \cvo{} film and the empty Ti $d$ bands of the \sto{} substrate. Thus, the main effect is the large tensile epitaxial strain that can be imposed on the \cvo{} film. Previous work showed that such a large epitaxial strain is expected to induce a metal-to-insulator transition in \cvo{}~\cite{Beck_et_al:2018}. Furthermore, it also leads to a strong orbital polarization, with fully occupied in-plane $d_{xy}$ orbitals and essentially empty $d_{xz}$ and $d_{yz}$. Consequently, confinement effects due to the reduced film thickness along $c$ are rather weak, and only noticeable in ultra-thin films of a few unit cells. Compared to the free-standing case analyzed in Ref.~\onlinecite{Beck_et_al:2018}, the presence of the \sto{} substrate further weakens these finite size effects, since the strong crystal-field splitting observed in the surface layer of the free-standing films~\cite{Beck_et_al:2018} is significantly reduced in the \cvo{}/\sto{} interface layer.
Thus, our results for the \cvo{}/\sto{} multilayers indicate that in some cases materials with vastly different properties can be sandwiched even in ultra-thin multilayers without substantial influence on the film itself, apart from the large effect of epitaxial strain.

In contrast to this, for the case of the \cvo{}/\lao{} heterostructures, the properties of the \cvo{} thin film can be altered significantly by the presence of the substrate, and even depend fundamentally on the specific electrostatic boundary conditions at the interface.
This follows from the different results obtained for different unit cell geometries and also from the different behavior at $p$-type compared to $n$-type interface in certain cases.

The weakest interface effect is observed in the symmetric multilayer slabs separated by vacuum, which thus partly resemble the case with the \sto{} substrate, albeit with a much lower tensile strain. Here, all V sites can assume a bulklike formal valence, and the gradient of the electrostatic potential inside the \lao{} layers does not seem to affect the \cvo{} layers.
This leaves the film properties independent of the interface termination and overall essentially bulklike, with only the film thickness reducing the effective dimensionality and thus supporting the metal-insulator transition for ultra-thin films of two unit cells.

On the other hand, strong interface effects can be observed for both symmetric and asymmetric multilayer geometries. Both cases exhibit a pronounced depletion or excess filling of the V-\ttg{} states at the $p$-type and $n$-type interfaces, respectively, even though the origin of this doping appears slightly different in the two cases (charge neutrality/stoichiometry of the periodic supercell in the case of the symmetric multilayers versus the internal field inside the asymmetric multilayers). Nevertheless, in both cases the doping is restricted to essentially one layer at the $p$-type interface, while it penetrates deeper into the \cvo{} at the $n$-type interface.
The interfacial doping naturally favors the metallic state of the \cvo{} film. However, finite size effects for a film thickness below 4 perovskite units, supported in some cases by an enhanced crystal field splitting close to the $p$-type interface, leads to an opposing tendency favoring the insulating state.

This competiton between doping and finite size or crystal-field effects leads to a non-monotonous thickness dependence for the asymmetric layers, strongly favoring the insulating state for only 1 or 2 layers of \cvo{} but more metallic compared to bulk for 3 or more layers, and nearly no change of \umit{} compared to bulk in the symmetric multilayers with $p$-type interfaces.
For the symmetric multilayers with the $n$-type interface, on the other hand, all layers are completely metallic, and there is no MIT at any reasonable $U$ value due to the weak crystal-field splitting and the more delocalized distribution of the excess charge (at least for the thicknesses considered in this work).

The different results obtained for the three different supercell geometries illustrate the strong sensitivity of the properties of polar interfaces from the specific electrostatic boundary conditions and the details of the charge compensation, even at chemically identical interfaces. 
Thus, calculations done for only one cell type have to be interpreted with care, and considering the specific stoichiometry and boundary conditions imposed by the symmetry of the cell.
From a different perspective, our results also suggest that if the boundary conditions at the interface can be controlled, or even specifically tuned, within an experimental setup, this can give access to an enlarged spectrum of materials properties though multilayer engineering. 
Of course in practice the precise control of the electrostatic boundary conditions might be challenging. Furthermore, other mechanisms of charge compensation at polar interfaces, not considered in our work, can occur, such as, e.g., cation interdiffusion or the formation of defects, in particular oxygen vacancies.

Finally, we note that in all the cases considered in this work, the mismatch in the octahedral tilt and rotation angles, or even in the bulk tilt systems of the two compounds, is accommodated within 1 or 2 layers away from the interface. In particular, the octahedral rotations around the $c$ direction, i.e. around the stacking direction of the mulitlayers, are barely affected at all by the presence of the interface. 
This suggests that interface engineering through the connectivity of octahedral tilts and rotations across an interface~\cite{Rondinelli/May/Freeland:2012} either requires a more pronounced difference between the bulk tilt angles than for the materials considered here or that it is more applicable to very short-period multilayers.
Interestingly however, it seems that the in-phase or out-of-phase coupling of the in-plane rotations across the \cvo{}/\lao{} interface is determined by the linking $A$-site cation, i.e., out-of-phase across the LaO/$n$-type interface and in-phase across the CaO/$p$-type interface. It will be interesting to see whether similar behavior can also be observed for other materials combinations.

In conclusion, we report on a large tunability of thin film properties as a function of
electrostatic boundary conditions in multilayer systems involving a polar discontinuity. The example of \cvo{} shows how the interplay between strain, finite size, interfacial doping, or the choice of substrate can significantly alter the competition between the metallic and insulating states in a ``Mott material''. Furthermore, it becomes clear that a comparison between experimental and computational results requires a careful consideration of the respective boundary conditions.

%%%%%%%%%%%%%%%%%%%%%%%%%%
\appendix*

\begin{acknowledgments}

We are grateful to Alexander Hampel for helpful discussions and support with technical aspects of the DMFT calculations.
This work was supported by ETH Zurich and the Swiss National Science Foundation through NCCR-MARVEL.
Calculations were performed on the cluster ``Piz Daint'' hosted by the Swiss National Supercomputing Centre.

\end{acknowledgments}

\bibliography{bibfile}

\end{document}